\documentclass[12pt,reqno]{amsart}

\usepackage{epsfig}
\usepackage{amsmath, amsthm, amsfonts}
\usepackage{graphicx}

\numberwithin{equation}{section}

\newcommand{\R}{{\mathbb R}}

\newcommand{\N}{{\mathbb N}}

\newcommand{\Z}{{\mathbb Z}}

\newcommand{\Tr}{{{\operatorname{Tr}}}}

\newcommand{\al}{\alpha}
\newcommand{\be}{\beta}
\newcommand{\ga}{\gamma}

\newcommand{\la}{\lambda}
\newcommand{\ep}{\varepsilon}
\newcommand{\de}{\delta}
\newcommand{\De}{\Delta}

\newcommand{\sg}{\sigma}
\newcommand{\om}{\omega}

\newcommand{\z}{\zeta}

\newtheorem{theo}{{\sc \bf Theorem}}[section]

\begin{document}

\title[Riemann-Hilbert Approach to the Six-Vertex Model]
{Riemann-Hilbert Approach to the Six-Vertex Model}

\author{Pavel Bleher}
\address{Department of Mathematical Sciences,
Indiana University-Purdue University Indianapolis,
402 N. Blackford St., Indianapolis, IN 46202, U.S.A.}
\email{bleher@math.iupui.edu}

\author{Karl Liechty}
\address{Department of Mathematics,
University of Michigan,
530 Church St., Ann Arbor, MI 48109, U.S.A.}
\email{kliechty@umich.edu}
\thanks{The first author is supported in part
by the National Science Foundation (NSF) Grant DMS-0969254.}

\date{\today}

\begin{abstract}
The six-vertex model, or the square ice model, with domain wall boundary conditions (DWBC)
has been introduced and solved for finite $n$ by Korepin and Izergin. The solution is based on
the Yang-Baxter equations and it represents the free energy in terms of an $n\times n$ Hankel 
determinant. Paul Zinn-Justin observed that the Izergin-Korepin formula can be re-expressed 
in terms of the partition function of a random matrix model with a nonpolynomial interaction. 
We use this observation to obtain the large $n$ asymptotics of the six-vertex model with DWBC.
The solution is based on the Riemann-Hilbert approach. In this paper we review asymptotic results
obtained in different regions of the phase diagram.
\end{abstract}

\maketitle

\section{Six-vertex model}

The {\it six-vertex model}, or the model of {\it two-dimensional ice}, is stated on a square 
lattice with arrows on edges. The arrows obey the rule that at every vertex there 
are two arrows 
%%%%%%%%%%%% Fig.  %%%%%%%%%%%%%%
\begin{center}
 \begin{figure}[h]\label{arrows}
\begin{center}
   \scalebox{0.52}{\includegraphics{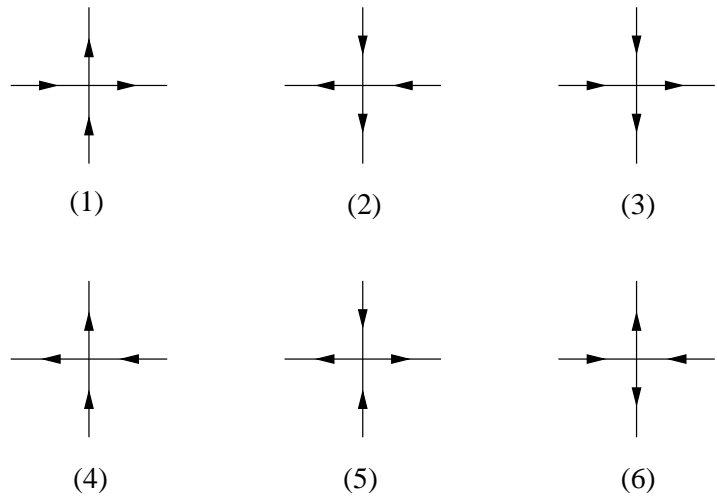}}
\end{center}
        \caption[The six arrow configurations allowed at a vertex]{The six arrow configurations allowed at a vertex.}
    \end{figure}
\end{center}
%%%%%%%%%%%%%%%%%%%%%%%%%%%%%%%%%%
pointing in and two arrows pointing out. This rule is sometimes 
called the {\rm ice-rule}. There are only six possible configurations of arrows at each 
vertex, hence the name of the model, see Fig.~1.

We will consider the {\it domain wall boundary conditions} (DWBC), 
in which the arrows on the upper and lower boundaries point into the square, 
and the ones on the left and right boundaries point out. 
One possible configuration with DWBC on the $4\times 4$ lattice is shown on Fig.~2.

%%%%%%%%%%%% Fig.  %%%%%%%%%%%%%%
\begin{center}
 \begin{figure}[h]\label{example1}
\begin{center}
   \scalebox{0.52}{\includegraphics{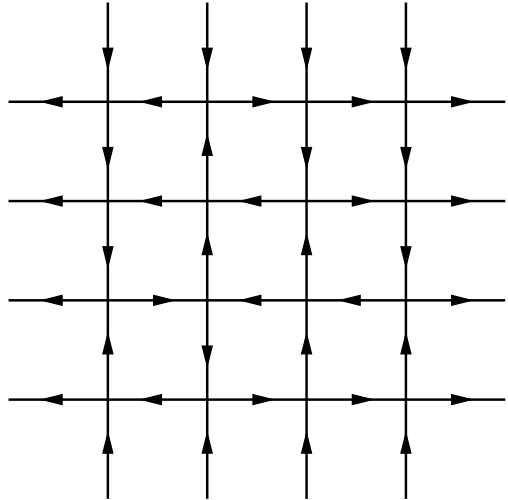}}
\end{center}
        \caption[An example of $4\times4$ configuration]{An example of $4\times4$ configuration.}
    \end{figure}
\end{center}
%%%%%%%%%%%%%%%%%%%%%%%%%%%%%%%%%%

The name of the {\it square ice} comes from the two-dimensional arrangement 
of water molecules, $H_2O$, with oxygen atoms at the vertices of 
the lattice and one hydrogen atom between each pair of adjacent oxygen 
atoms. We place an arrow in the direction from  a hydrogen atom 
toward an oxygen atom if there is a bond between them. Thus, as 
we already noticed before, there are two in-bound and two out-bound 
arrows at each vertex.

%%%%%%%%%%%% Fig. %%%%%%%%%%%%%%
\begin{center}
 \begin{figure}[h]\label{ice1}
\begin{center}
   \scalebox{0.52}{\includegraphics{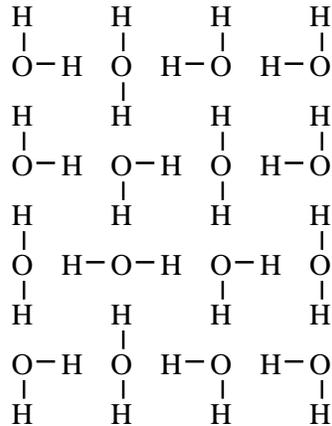}}
\end{center}
        \caption[The corresponding ice model]{The corresponding ice model.}
    \end{figure}
\end{center}
%%%%%%%%%%%%%%%%%%%%%%%%%%%%%%%%%%

For each possible vertex state we assign a {\it weight} $w_i,\; i=1,\dots,6$, 
and define, as usual, the {\it partition function}, as a sum over all possible 
arrow configurations of the product of the vertex weights,
\begin{equation}\label{lattice_11}
Z_n=\sum_{{\rm arrow\; configurations}\;\sigma}w(\sigma),
\qquad w(\sigma)=\prod_{x\in V_n} w_{\sg(x)}=\prod_{i=1}^6w_i^{N_i(\sigma)},
\end{equation}
where $V_n$ is the $n\times n$ set of vertices,
$\sg(x)\in\{1,\ldots,6\}$ is the vertex configuration of $\sg$
at vertex $x$, according to Fig. 1, and
$N_i(\sigma)$ is the number of vertices of  type $i$ in  
the configuration $\sg$. The sum is taken over all possible configurations
obeying the given boundary condition. The {\it Gibbs measure} is defined then
as
\begin{equation}\label{lattice_12}
\mu_n(\sg)=\frac{w(\sg)}{Z_n}\,.
\end{equation}
Our main goal is to obtain the {\it large $n$ asymptotics} of the partition function $Z_n$.

 In general, the six-vertex model has {\it six parameters}: the weights $w_i$.  
However, by using some conservation laws we can reduce these to only {\it two parameters}.  Any fixed boundary conditions impose some conservation laws on the six-vertex model.  In the case of DWBC, they are
\begin{equation}\label{par1}
N_1(\sg)=N_2(\sg), \quad N_3(\sg)=N_4(\sg),\quad N_5(\sg)=N_6(\sg)+n.
\end{equation}
This allows us to reduce to the case 
\begin{equation}\label{par}
w_1=w_2\equiv a, \quad
w_3=w_4\equiv b, \quad
w_5=w_6\equiv c.
 \end{equation}
Then by using the identity,
\begin{equation}\label{cl_17}
Z_n(a,a,b,b,c,c)=c^{n^2}Z_n\left(\frac{a}{c},\frac{a}{c},\frac{b}{c},\frac{b}{c},1,1\right),
\end{equation}
 we can reduce to the two parameters, $\frac{a}{c}$ and $\frac{b}{c}\,$.
For details on how we make this reduction, see, e.g., the works \cite{AR} of Allison and Reshetikhin,
\cite{FS} of Ferrari and Spohn, and \cite{BL1} of Bleher and Liechty. 

\section {Phase diagram of the six-vertex model}

Introduce the parameter
\begin{equation}\label{pf1}
\Delta=\frac{a^2+b^2-c^2}{2ab}\,.
\end{equation}
The {\it phase diagram} of the six-vertex model consists of the following three
 regions: the {\it ferroelectric phase region}, $\Delta > 1$; the {\it anti-ferroelectric phase region}, 
$\Delta<-1$; and, the {\it disordered phase region}, $-1<\Delta<1$, see, e.g., \cite{LW}. 
In these three regions we parameterize the weights in the standard way:
in the ferroelectric phase region,
\begin{equation}\label{pf4}
a=\sinh(t-\ga), \quad
b=\sinh(t+\ga), \quad
c=\sinh(2|\ga|), \quad
0<|\ga|<t,
\end{equation}
in the anti-ferroelectric phase region,
\begin{equation}\label{pf5}
a=\sinh(\ga-t), \quad
b=\sinh(\ga+t), \quad
c=\sinh(2\ga), \quad
|t|<\ga,
\end{equation}
and in the disordered phase region,
\begin{equation}\label{pf6}
a=\sin(\ga-t), \quad
b=\sin(\ga+t), \quad
c=\sin(2\ga), \quad
|t|<\ga.
\end{equation}
The phase diagram of the model is shown on Fig.~4.

%%%%%%%%%%%% Fig.  %%%%%%%%%%%%%%
\begin{center}
 \begin{figure}[h]\label{PhaseDiagram1}
\begin{center}
   \scalebox{0.6}{\includegraphics{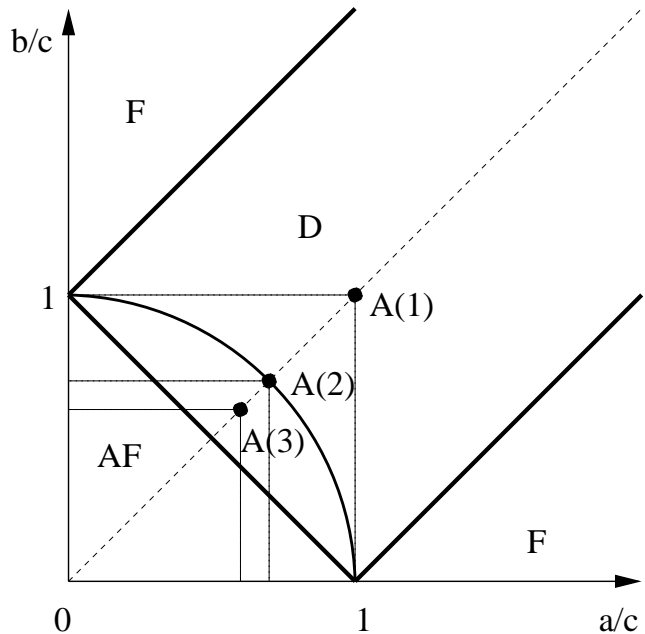}}
\end{center}
        \caption[The phase diagram of the model]{The phase diagram of the model,
 where {\bf F}, {\bf AF} and {\bf D} mark ferroelectric, antiferroelectric,  
and disordered  phases, respectively. The circular arc corresponds to the 
so-called ``free fermion'' line, where $\Delta=0$, and the three
dots correspond to 1-, 2-, and 3-enumeration of alternating sign matrices.}
    \end{figure}
\end{center}
%%%%%%%%%%%%%%%%%%%%%%%%%%%%%%%%%%

The phase diagram  and the Bethe-Ansatz solution  of the {\it six-vertex model for periodic and anti-periodic
boundary conditions} are thoroughly
discussed in the works of Lieb \cite{Lieb1}-\cite{Lieb4}, Lieb, Wu \cite{LW},
Sutherland \cite{Sut}, Baxter \cite{Bax}, Batchelor, Baxter, O'Rourke, Yung \cite{BBOY}.
See also the work of Wu, Lin \cite{WL}, in which  the Pfaffian solution for the six-vertex
model with periodic boundary conditions is obtained on the free fermion line, $\Delta=0$.

\section{Izergin-Korepin determinantal formula}

 The {\it six-vertex model with DWBC} was introduced 
by Korepin in \cite{Kor}, who derived an important 
recursion relation for the partition function of the model.
This lead to a beautiful {\it determinantal formula} of Izergin and Korepin \cite{Ize},
for the partition function of the six-vertex model with DWBC.
A detailed proof of this formula and its generalizations are given in
the paper of Izergin, Coker, and Korepin \cite{ICK}. When the weights are 
parameterized according to (\ref{pf6}), the formula of Izergin-Korepin is
\begin{equation} \label{pf7}
Z_n=\frac{(ab)^{n^2}}{\left(
\prod_{k=0}^{n-1}k!\right)^2}\,\tau_n\,,
\end{equation}
where $\tau_n$ is the Hankel determinant,
\begin{equation} \label{pf8}
\tau_n=\det\left(\frac{d^{i+k-2}\phi}{dt^{i+k-2}}\right)_{1\le i,k\le n},
\end{equation} 
and
\begin{equation} \label{pf9}
\phi(t)=\frac{c}{ab}\,.
\end{equation}
Observe that $a,b,c$ have different parameterizations \eqref{pf4}--\eqref{pf6} in different phase regions.
An elegant derivation of the 
Izergin-Korepin determinantal formula from the {\it Yang-Baxter equations} is 
given in the papers of Korepin and Zinn-Justin \cite{KZ} and Kuperberg \cite {Kup}.
 
One of the applications of the determinantal formula is that it
implies that 
the partition function $\tau_n$ solves the {\it Toda equation},
\begin{equation} \label{pf10}
\tau_N\tau''_n-{\tau'_n}^2=\tau_{n+1}\tau_{n-1},
\qquad n\ge 1,\qquad ({}')=\frac{\partial }{\partial t}\,,
\end{equation}
cf. \cite{Sog}. This was used by Korepin and Zinn-Justin \cite{KZ} to 
derive the free energy of the six-vertex model with DWBC, assuming
some Ansatz on the behavior of subdominant terms in the large $n$
asymptotics of the free energy.

\section{The six-vertex model with DWBC and a random matrix model}

Another application of the Izergin-Korepin determinantal formula is that
$\tau_n$ can be expressed in terms of  a partition function of
a {\it random matrix model}. The relation to the random matrix model was
obtained and used by Zinn-Justin \cite{Z-J1}. It can be 
derived as follows. Consider first the disordered phase region.

\subsection{Disordered phase region.}
For the evaluation of the Hankel determinant, it is convenient 
to use an integral representation of the function
\begin{equation} \label{dph6}
\phi(t)=\frac {\sin 2\ga}{\sin(\ga-t)\sin(\ga+t)}\,,
\end{equation} 
namely, to write it 
in the form of the Laplace transform,
\begin{equation} \label{dph6}
\phi(t)=\int_{-\infty}^\infty e^{t\la}m(\la)d\la,
\end{equation} 
where
\begin{equation} \label{dph7}
m(\la)=\frac{\sinh\frac{\la}{2}(\pi-2\ga)}
{\sinh\frac{\la}{2}\pi}\,.
\end{equation} 
Then
\begin{equation} \label{dph8}
\frac{d^i\phi}{dt^i}=\int_{-\infty}^\infty \la^i
e^{t\la}m(\la)d\la,
\end{equation} 
and by substituting this into the Hankel determinant,
(\ref{pf8}), we obtain that
\begin{equation} \label{dph9}
\begin{aligned}
\tau_n&=\int \prod_{i=1}^n[e^{t\la_i}m(\la_i)d\la_i]
\det(\la_i^{i+k-2})_{1\le i,k\le n}\\
&=\int \prod_{i=1}^n[e^{t\la_i}m(\la_i)d\la_i]
\det(\la_i^{k-1})_{1\le i,k\le n}\prod_{i=1}^n\la_i^{i-1}.
\end{aligned}
\end{equation}
Consider any permutation $\sg\in S_n$ of variables $\la_i$. 
From the last equation we have that
\begin{equation} \label{dph10}
\tau_n=\int \prod_{i=1}^n[e^{t\la_i}m(\la_i)d\la_i]
(-1)^\sg\det(\la_i^{k-1})_{1\le i,k\le n}\prod_{i=1}^n\la_{\sg(i)}^{i-1}.
\end{equation}
By summing over $\sg\in S_n$, we obtain that
\begin{equation} \label{dph11}
\tau_n=\frac{1}{n!}\int \prod_{i=1}^n[e^{t\la_i}m(\la_i)d\la_i]
\De(\la)^2,
\end{equation}
where $\De(\la)$ is the Vandermonde determinant,
\begin{equation} \label{dph12}
\De(\la)=\det(\la_i^{k-1})_{1\le i,k\le n}=\prod_{i<k}(\la_k-\la_i).
\end{equation}
Equation (\ref{dph11}) expresses $\tau_n$ in terms of a matrix model
integral. Namely, if $m(x)=e^{-V(x)}$, then
\begin{equation} \label{dph13}
\tau_n=\frac{\prod_{n=0}^{n-1}n!}{\pi^{n(n-1)/2}}\int dM e^{\Tr [tM-V(M)]},
\end{equation}
where the integration is over the space of $n\times n$ Hermitian matrices.
The matrix model integral can be solved, furthermore, in terms of
{\it orthogonal polynomials}.

Introduce monic polynomials $P_k(x)=x^k+\dots$ orthogonal 
on the line with respect to the weight 
\begin{equation} \label{dph13a}
w(x)=e^{tx}m(x),
\end{equation}
so that
\begin{equation} \label{dph14}
\int_{-\infty}^\infty P_j(x)P_k(x)e^{tx}m(x)dx=h_k\delta_{nm}.
\end{equation}
Then it follows from (\ref{dph11}) that
\begin{equation} \label{dph15}
\tau_n=\prod_{k=0}^{n-1}h_k. 
\end{equation}
The orthogonal polynomials satisfy 
the three term recurrence relation,
\begin{equation} \label{dph16}
xP_k(x)=P_{k+1}(x)+Q_kP_n(x)+R_kP_{k-1}(x),
\end{equation}
where $R_n$ can be found as
\begin{equation} \label{dph16a}
R_n=\frac{h_k}{h_{k-1}}\,,
\end{equation}
see, e.g., \cite{Sze}.
This gives that
\begin{equation} \label{dph17}
h_k=h_0\prod_{j=1}^k R_j,
\end{equation}
where
\begin{equation}\label{dph17a}
h_0=\int_{-\infty}^\infty e^{tx}m(x)dx
=\frac{\sin(2\ga)}{\sin(\ga+t)\sin(\ga-t)}\,.
\end{equation}
By substituting (\ref{dph17}) into (\ref{dph15}), we obtain that
\begin{equation} \label{dph18}
\tau_n=h_0^n\prod_{k=1}^{n-1}R_n^{n-k}. 
\end{equation}

\subsection{Ferroelectric phase} In the ferroelectric phase, the parameters $a, b$, and $c$ are parameterized by (\ref{pf4}). We consider the case $\ga>0$, which corresponds to the region $b>a+c$ in the phase diagram.  The case $\ga<0$ is similar, and $a$ and $b$ should be exchanged in that case.  The function $\phi$ is the Laplace transform of a discrete measure supported on the positive integers:
\begin{equation} \label{fe2}
\phi(t)=\frac{\sinh(2\ga)}{\sinh(t+\ga)\sinh(t-\ga)}=
4\sum_{l=1}^\infty e^{-2tl}\sinh(2\ga l).
\end{equation} 
Then, similar to (\ref{dph11}), we find that
\begin{equation} \label{fe3}
\tau_n=\frac{2^{n^2}}{n!}\sum_{l_1,\ldots,l_n=1}^\infty \Delta(l_i)^2\prod_{i=1}^n
\left[2e^{-2tl_i}\sinh(2\ga l_i)\right].
\end{equation}
This is the partition function for a discrete version of a Hermitian
random matrix model, often called a {\it discrete orthogonal polynomial ensemble} (DOPE), and can also be solved in terms of orthogonal polynomials.  The appropriate polynomials in this case are the monic polynomials $P_n(l)= l^n + \cdots$ with the orthogonality
\begin{equation} \label{fe4}
\sum_{l=1}^\infty P_j(l)P_k(l)w(l)=h_k\delta_{jk}\,, \quad w(l)=2e^{-2tl}\sinh(2\ga l)=e^{-2tl+2\ga l}-e^{-2tl-2\ga l}.
\end{equation}
Then it follows from (\ref{fe3}) that
\begin{equation} \label{fe5}
\tau_n=2^{n^2}\prod_{k=0}^{n-1}h_k.
\end{equation}

\subsection{Critical line between disordered and ferroelectric phase}
When the parameters $a, b$, and $c$ are such that $b-a=c$, (so $\De =1$ in (\ref{pf1})), the Izergin-Korepin formula is not directly applicable.  However, we may consider a limiting case of the orthogonal polynomial formula (\ref{fe5}).  On the critical line
\begin{equation} \label{cr1}
\frac{b}{c}-\frac{a}{c}=1,
\end{equation}
we fix a point,
\begin{equation} \label{cr2}
\frac{a}{c}=\frac{\al-1}{2}\,,\quad \frac{b}{c}=\frac{\al+1}{2}\,;\qquad \al>1,
\end{equation}
and consider the partition function 
\begin{equation} \label{cr3}
Z_n=Z_n\left(\frac{\al-1}{2}\,,\frac{\al-1}{2}\,,\frac{\al+1}{2}\,,\frac{\al+1}{2}\,,1,1\right).
\end{equation}
Consider
the limit of (\ref{fe5}) as
\begin{equation} \label{cr4}
t,\ga\to +0,\qquad \frac{t}{\ga}\to \al.
\end{equation}
Observe that in this limit,
\begin{equation} \label{cr5}
\frac{a}{c}=\frac{\sinh(t-\ga)}{\sinh(2\ga)}\to \frac{\al-1}{2},\qquad 
\frac{b}{c}=\frac{\sinh(t+\ga)}{\sinh(2\ga)}\to \frac{\al+1}{2}.
\end{equation}
By (\ref{cl_17}),  (\ref{pf7}), and (\ref{dph15}), we have
\begin{equation} \label{cr6}
Z_n\left(\frac{a}{c},\frac{a}{c},\frac{b}{c},\frac{b}{c},1,1\right)=
\left[\frac{2\sinh(t-\ga)\sinh(t+\ga)}{\sinh(2\ga)}\right]^{n^2}\prod_{k=0}^{n-1}\frac{h_k}{(k!)^2}\,. 
\end{equation}
To deal with limit (\ref{cr4}) we  need to rescale the orthogonal polynomials $P_k(l)$. 
Introduce the rescaled variable,
\begin{equation} \label{cr7}
x=2tl-2\ga l,
\end{equation}
and the rescaled limiting weight,
\begin{equation} \label{cr8}
w_{\al}(x)=\lim_{t,\ga\to +0,\; \frac{t}{\ga}\to \al} (e^{-2tl+2\ga l}-e^{-2tl-2\ga l})=e^{-x}-e^{-rx},
\qquad r=\frac{\al+1}{\al-1}>1\,.
\end{equation}
Consider monic orthogonal polynomials $P_j(x;\al)$ satisfying the orthogonality condition,
\begin{equation} \label{cr9}
\int_0^\infty P_j(x;\al)P_k(x;\al)w_{\al}(x)dx=h_{k,\al}\de_{jk}\,.
\end{equation}
To find a relation between $P_k(l)$ and $P_k(x;\al)$, introduce the monic polynomials
\begin{equation} \label{cr9a}
\tilde P_k(x)=\de^kP_k(x/\de),
\end{equation}
where
\begin{equation} \label{cr10}
\de=2t-2\ga,
\end{equation}
and rewrite orthogonality
condition (\ref{dph14}) in the form 
\begin{equation} \label{cr11}
\sum_{l=1}^\infty \tilde P_j(l\de)\tilde P_k(l\de)w_\al(l\de)\de=\de^{2k+1}h_k\delta_{jk},
\end{equation}
which is a Riemann sum for the integral in orthogonality condition (\ref{cr9}). Therefore,
\begin{equation} \label{cr12}
\lim_{t,\ga\to +0,\; \frac{t}{\ga}\to \al}\tilde P_k(x)=P_k(x;\al), \quad
\lim_{t,\ga\to +0,\; \frac{t}{\ga}\to \al}\de^{2k+1}h_k=h_{k,\al}.
\end{equation}
Thus, if we rewrite formula (\ref{cr6}) as
\begin{equation} \label{cr14}
Z_n\left(\frac{a}{c},\frac{a}{c},\frac{b}{c},\frac{b}{c},1,1\right)=
\left[\frac{2\sinh(t-\ga)\sinh(t+\ga)}{\sinh(2\ga)\de}\right]^{n^2}\prod_{k=0}^{n-1}\frac{\de^{2k+1}h_k}{(k!)^2}\,, 
\end{equation}
we can take limit (\ref{cr4}). In the limit we obtain that
\begin{equation} \label{cr15}
Z_n=Z_n\left(\frac{\al-1}{2}\,,\frac{\al-1}{2}\,,\frac{\al+1}{2}\,,\frac{\al+1}{2}\,,1,1\right)=
\left(\frac{\al+1}{2}\right)^{n^2}\prod_{k=0}^{n-1}\frac{h_{k,\al}}{(k!)^2}\,.
\end{equation}

\subsection{Antiferroelectric phase} 
In the antiferroelectric phase, the parameters $a, b$, and $c$ are parameterized by (\ref{pf5}), and the function 
\begin{equation}\label{af1}
\phi(t)= \frac{\sinh (2\ga)}{\sinh(\ga -t)\sinh(\ga+t)}\,, \quad |t|<\ga,
\end{equation}
is the Laplace transform of a discrete measure supported on the integers:
\begin{equation} \label{af2}
\phi(t)=\frac{\sinh(2\ga)}{\sinh(\ga-t)\sinh(\ga+t)}=
2\sum_{l=-\infty}^\infty e^{2tl-2\ga |l|}.
\end{equation} 
Then
\begin{equation} \label{af3}
\tau_n=\frac{2^{n^2}}{n!}\sum_{l_1,\ldots,l_n=-\infty}^\infty \De(l)^2\prod_{i=1}^n
e^{2tl_i-2\ga |l_i|}.
\end{equation}
This is again the partition function of a DOPE, and we introduce the discrete monic polynomials $P_n(l)=l^n+\dots$ via the orthogonality condition
\begin{equation} \label{af4}
\sum_{l=-\infty}^\infty P_j(l)P_k(l)w(l)=h_k\delta_{jk}\,, \quad w(l)=e^{2tl-2\ga |l|}.
\end{equation}
Then it follows from (\ref{af3}) that
\begin{equation} \label{af5}
\tau_n=2^{n^2}\prod_{k=0}^{n-1}h_k.
\end{equation}

\subsection{Critical line between the antiferroelectric and disordered phases}
When the parameters $a, b$, and $c$ are such that $a+b=c$, (so $\De =-1$ in (\ref{pf1})), the Izergin-Korepin formula is not directly applicable, and we must consider a limiting case of the orthogonal polynomial formula (\ref{af5}).  On the critical line
\begin{equation} \label{afd1}
\frac{a}{c}+\frac{b}{c}=1,
\end{equation}
we fix a point,
\begin{equation} \label{afd2}
\frac{a}{c}=\frac{1-\al}{2}\,,\quad \frac{b}{c}=\frac{1+\al}{2}\,,\qquad -1<\al<1,
\end{equation}
and consider the partition function 
\begin{equation} \label{afd3}
Z_n=Z_n\left(\frac{1-\al}{2}\,,\frac{1-\al}{2}\,,\frac{1+\al}{2}\,,\frac{1+\al}{2}\,,1,1\right).
\end{equation}
This corresponds to taking a limit of the Izergin-Korepin formula in the antiferroelectric phase as $t, \ga \to 0$, and $t/\ga =\al$.
Introduce the rescaled variable,
\begin{equation} \label{afd4}
x=-2tl+2\ga l,
\end{equation}
and the rescaled limiting weight,
\begin{equation} \label{afd5}
w_{\al}(x)=\lim_{t,\ga\to +0,\; \frac{t}{\ga}\to \al} e^{2tl-2\ga |l|}=
\left\{
\begin{aligned}
&e^{-x}\,, \quad x \ge 0 \\
&e^{rx} \,, \quad x <0\,,
\end{aligned}
\right.
\end{equation}
where
\begin{equation}
\qquad r=\frac{1+\al}{1-\al}>0\,.
\end{equation}
Consider monic orthogonal polynomials $P_j(x;\al)$ satisfying the orthogonality condition,
\begin{equation} \label{afd6}
\int_\R P_j(x;\al)P_k(x;\al)w_{\al}(x)dx=h_{k,\al}\de_{jk}\,,
\end{equation}
which can be obtained from the polynomials (\ref{af4}) by taking the appropriate scaling limit as $t, \ga \to 0$, and $t/\ga =\al$.  Similar to (\ref{cr15}), we obtain
\begin{equation} \label{afd7}
Z_n=Z_n\left(\frac{\al-1}{2}\,,\frac{\al-1}{2}\,,\frac{\al+1}{2}\,,\frac{\al+1}{2}\,,1,1\right)=
\left(\frac{1+\al}{2}\right)^{n^2}\prod_{k=0}^{n-1}\frac{h_{k,\al}}{(k!)^2}\,.
\end{equation}

\section{Large $n$ asymptotics of $Z_n$}

The asymptotic evaluation of $Z_n$ in the different regions of the phase diagram thus reduces to asymptotic evaluation of different systems of orthogonal polynomials.  In general, this may be done by formulating the orthogonal polynomials as the solution to a $ 2\times 2$ matrix valued Riemann-Hilbert problem as in \cite{FIK}.  One may then perform the steepest descent analysis of Deift and Zhou \cite{DZ}.  In the case that the weight of orthogonality is a continuous one on $\R$, this analysis was performed for weights of the form $\exp(-nV(x))$ for a very general class of analytic potential functions $V(x)$ in 
\cite{DKMVZ}.   The analysis was adapted to the case that the orthogonality is with respect to a discrete measure in \cite{BKMM} and \cite{BL4}.  The steepest descent analysis yields the following results in the different regions of the phase diagram.

\subsection{Disordered phase}

\begin{theo} \label{kappa} (See \cite{BF}.)
 Let the weights $a$, $b$, and $c$, in the six-vertex model with DWBC be parameterized as in (\ref{pf6}).  Then, as $n \to \infty$, the partition function $Z_n$ has the asymptotic expansion
\begin{equation} \label{dph28}
Z_n= Cn^{\kappa}F^{n^2}\left(1+O(n^{-\ep})\right),\qquad \ep>0,
\end{equation}
where
\begin{equation} \label{dph29}
F=\frac{\pi ab}{2\ga \cos\left(\frac{\pi t}{2\ga}\right)}\,, \quad \kappa=\frac{1}{12}-\frac{2\ga^2}{3\pi(\pi-2\ga)}\,,
\end{equation}
and $C>0$ is a constant.
\end{theo} 

This proves the conjecture of Zinn-Justin, and it gives the exact 
value of the exponent $\kappa$. Let us remark, that the presence of the power-like factor
$n^\kappa$ in the asymptotic expansion of $Z_n$ in (\ref{dph28})
is rather unusual from the point of view of random matrix models. Also,
in the one-cut case the usual large $n$ asymptotics of $Z_n$ in a non-critical 
random matrix model is the so called ``topological expansion'', which gives $Z_n$ as an asymptotic 
series in powers of $1/n^2$ (see e.g.
\cite{EM} and \cite{BI3}). In this case the asymptotic expansion goes over, in general, non-integer 
inverse powers of $n$ (see \cite{BF}). 

It is noteworthy that, as shown in \cite{BKZ}, asymptotic formula \eqref{dph28} remains valid on the borderline between  
the disordered and antiferroelectric phases. In this case $\kappa=\frac{1}{12}\,$, which
corresponds to $\ga=0$.

\subsection{Ferroelectric phase}

Bleher and Liechty \cite{BL1}, \cite{BL2} obtained the large $n$ asymptotics of $Z_n$ 
in the ferroelectric phase, $\De>1$,  
and also on the critical line between the ferroelectric and disordered phases, $\De=1$. In the
ferroelectric phase we use parameterization (\ref{pf4}) for $a,b$ and $c$. The large $n$ asymptotics 
of $Z_n$ in the ferroelectric phase is given by the following theorem.

\begin{theo} \label{fe} (See \cite{BL1}.)
 Let the weights $a$, $b$, and $c$ in the six-vertex model with DWBC be parameterized as in (\ref{pf4}) with $t>\ga>0$.  
For any $\ep>0$, as $n\to\infty$,
\begin{equation} \label{dph30}
Z_n=C G^n F^{n^2}\left[1+O\left(e^{- n^{1-\ep}}\right)\right],
\end{equation}
where $C=1-e^{-4\ga}$, $G=e^{\ga-t}$, and $F=b$.
\end{theo}

On the critical line  between the ferroelectric and disordered phases we use the parameterization
$b=a+1$, $c=1$. The main result here is the following asymptotic formula for $Z_n$.

\begin{theo}\label{fcr} (See \cite{BL2}.)
As $n\to\infty$,
\begin{equation} \label{main7} 
Z_n=C n^\kappa
G^{\sqrt n}F^{n^2}[1+O(n^{-1/2})]\,,
\end{equation}
where $C>0$,
\begin{equation} \label{main8}
\kappa=\frac{1}{4}\,,\qquad G=\exp\left[-\zeta\left(\frac{3}{2}\right)\sqrt{\frac{a}{ \pi }}\right]\,,
\end{equation}
and
\begin{equation} \label{main9}
F=b\,.
\end{equation}
\end{theo}

Notice that in both Theorem \ref{fe} and in Theorem \ref{fcr}, the limiting free energy $F$ is the weight $b$.  The ground state in this phase is unique and is achieved when there is exactly one $c$-type vertex in each row and column, and the rest of the vertices are of type $b$.  That is, the diagonal consists of type $5$ vertices while above the diagonal all vertices are type $3$ and below all vertices are type $4$.  The weight of the ground state is $b^{n^2} (c/b)^{n}$, and thus the free energy in the ferroelectric phase is completely determined by the ground state.  This is a reflection of the fact that local fluctuations from the ground state can take place only in a thin neighborhood of the diagonal.  The conservation laws (\ref{par1}) forbid local fluctuations away from the diagonal.

\subsection{Antiferroelectric phase}
The large $n$ asymptotics in the antiferroelectric phase were obtained by Bleher and Liechty in \cite{BL3}.  They are given in the following theorem.  In this theorem $\vartheta_1$ and $\vartheta_4$ are the Jacobi theta functions with elliptic nome $q=e^{-\pi^2/2\ga}$ (see e.g., \cite{WW}), and the phase $\om$ is given as
\begin{equation}
\om=\frac{\pi}{2} \left(1+\frac{t}{\ga}\right).
\end{equation}

\begin{theo} \label{thmain1}  (See \cite{BL3}.)
Let the weights $a$, $b$, and $c$ in the six-vertex model with DWBC be parameterized as in (\ref{pf4}).  As $n\to\infty$, 
\begin{equation}\label{main17}
Z_n=C\vartheta_4\left(n\om\right) F^{n^2}(1+O(n^{-1})),
\end{equation}
where $C>0$ is a constant, and
\begin{equation}\label{main18}
F=\frac{\pi ab \vartheta'_1(0)}{2\ga\vartheta_1(\om)}\,.
\end{equation}
\end{theo}

In contrast to the disordered phase, note the lack of a power like term.  In contrast to the ferroelectric phase, notice that the free energy depends transcendentally on the weight of the ground state configuration.  Only in the limit as $\ga \to \infty$, which can be regarded as the low temperature limit, does the the weight of the ground state become dominant.  For a discussion of this limit, see \cite{Z-J1}.

\section{The Riemann-Hilbert approach}

All the above asymptotic results are obtained in the Riemann-Hilbert approach, but the concrete 
asymptotic analysis of the Riemann-Hilbert problem is quite different in the different phase regions. Let us discuss it.

\subsection{Disordered phase region} To apply the Riemann-Hilbert approach, we introduce a rescaled
weight as 
\begin{equation}\label{rw1}
w_n(x)=w\left(\frac{nx}{\ga}\right)\,.
\end{equation}
It can be written as 
\begin{equation}\label{rw2}
w_n(x)=e^{-nV_n(x)},
\end{equation}
where
\begin{equation} \label{rw3}
V_{n}(x)=-\z x-\frac{1}{n}\,\ln\frac{
\sinh \left(n\left(\frac{\pi}{2\ga}-1\right)x\right)}
{\sinh\left(\frac{n\pi x}{2\ga}\right)}\,,\qquad \z=\frac{t}{\ga}\,.
\end{equation}
The external potential $V_n(x)$ is real analytic for any finite $n$, but it has logarithmic 
singularities on the imaginary axes, which accumulate to the origin as $n\to \infty$. In fact,
the limiting external potential,
\begin{equation} \label{rw4}
 \lim_{n\to\infty}V_n(x)= V(x) =
-\zeta x+|x|,
\end{equation}
is not analytic at $x=0$. The Riemann-Hilbert approach developed in \cite{BF} is based on opening of
lenses whose boundary approaches the origin as $n\to\infty$. This turns out to be possible due to the fact
that the density of the equilibrium measure $\rho_n(x)$ for the external potential $V_n(x)$ diverges
logarithmically at the origin as $n\to\infty$, and as a result, the jump matrix 
on the boundary of the lenses converges to the unit matrix  (for details see  \cite{BF}).
The calculation of subdominant asymptotic terms in the partition function as $n\to\infty$ is the
central difficult part of the work \cite{BF}, and it is done by an asymptotic analysis of the solution
to the Riemann-Hilbert problem near the turning points and near the origin.

\subsection{Ferroelectric phase region} 
In the ferroelectric region, the measure of orthogonality is a discrete one on $\N$.  To apply the Riemann-Hilbert approach to discrete orthogonal polynomials, we need to rescale both the weight and the lattice that supports the measure so that the mesh of the lattice goes to zero as $n \to \infty$.  Introduce the rescaled lattice and weight
\begin{equation}\label{rhf1}
L_n=\left(\frac{2t}{n}\right) \N\,, \qquad w_n(x)=e^{-nx(1-\z)} \left(1-e^{-4nx}\right)=e^{-nV_n(x)}\,,
\end{equation}
where
\begin{equation}\label{rhf2}
V_n(x)=x(1-\z)-\frac{1}{n}\log\left(1-e^{-2nx\z}\right)\,, \qquad 0<\z=\frac{\ga}{t} <1.
\end{equation}
Then the orthogonality condition (\ref{fe4}) can be written as
\begin{equation}\label{rhf3}
\sum_{x\in L_n} P_j\left(\frac{nx}{2t}\right) P_k\left(\frac{nx}{2t}\right) w_n(x) = h_k \de_{jk}.
\end{equation}
Notice that, as $n\to \infty$, $V_n(x)$ has the limit
\begin{equation}\label{rhf4}
\lim_{n \to \infty} V_n(x)=x(1-\z), 
\end{equation}
which would indicate that, in the large $n$ limit, the polynomials (\ref{fe4}) behave as polynomials orthogonal on $\N$ with a simple exponential weight.  These polynomials are a special case of the classical {\it Meixner polynomials}, and there are exact formulae for their recurrence coefficients (see e.g., \cite{KS}).
The monic Meixner polynomials which concern us are defined from the orthogonality condition
\begin{equation} \label{rhf5}
\sum_{l=1}^\infty Q_j(l)Q_k(l)q^l=h_k^{\rm Q}\de_{jk},
\qquad  \quad q=e^{2\ga-2t}\,,
\end{equation} 
and the normalizing constants are given exactly as
\begin{equation}\label{rhf6}
h_k^{\rm Q}= \frac{(k!)^2q^{k+1}}{(1-q)^{2k+1}}\,.
\end{equation}
Up to the constant factor, Theorem \ref{fe} can therefore be proven by showing that $h_k$ and $h_k^{\rm Q}$ are asymptotically close as $n\to \infty$.  More precisely, it is shown in \cite{BL1} that as $k \to \infty$, for any $\ep >0$,
\begin{equation}\label{rhf7}
h_k=h_k^{\rm Q}\left(1+O\left(e^{-k^{1-\ep}}\right)\right).
\end{equation}

\subsection{Antiferroelectric region}
In the antiferroelectric region, the orthogonal polynomials are with respect to a discrete weight, and we rescale the weight in (\ref{af4}) and the integer lattice as
\begin{equation}\label{rha1}
L_n=\left(\frac{2\ga}{n}\right) \Z\,, \qquad w_n(x)=e^{-nV(x)}\,, \quad V(x)=|x|-\z x\,, \quad \z=\frac{t}{\ga} <1,
\end{equation}
so that the orthogonality condition (\ref{af4}) can be written as
\begin{equation}\label{rha2}
\sum_{x\in L_n} P_j\left(\frac{nx}{2\ga}\right)P_k\left(\frac{nx}{2\ga}\right) w_n(x) = h_k \de_{jk}.
\end{equation}

The mesh of the lattice $L_n$ is $2\ga/n$, which places an upper constraint on the equilibrium measure, which is the limiting distribution of zeroes of the orthogonal polynomials.  This upper constraint is realized. The equilibrium measure, which has density $\rho(x)$, is supported on a single interval $[\al, \be]$, but within that interval is an interval $[\al', \be']$ on which $\rho(x) \equiv 1/2\ga$.  This interval is called the {\it saturated region}, and it separates the single band of support $[\al, \be]$ into the two analytic bands $[\al, \al']$ and $[\be', \be]$.  Thus in effect we have a ``two-cut" situation, which is the source of the quasi-periodic factor $\vartheta_4(n \om)$ in Theorem \ref{thmain1}.  

In principle, a problem could come from the fact that the potential $V(x)$ is not analytic at the origin.  However, it turns out that this point of nonanalycity is always in the saturated region and therefore does not present a problem in the steepest descent analysis.

As previously noted, there is no power-like term in the asymptotic formula for $Z_n$ in the antiferroelectric phase.  The Riemann-Hilbert approach to orthogonal polynomials generally gives an expansion of the normalizing constants $h_n$ in inverse powers of $n$.  In the two-cut case, the coefficients in this expansion may be quasi-periodic functions of $n$. For the orthogonal polynomials (\ref{af4}) it is a tedious calculation involving the Jacobi theta functions to show that the term of order $n^{-1}$ vanishes in the expansion of $h_n$, which then implies the absence of the power-like term in $Z_n$.


\begin{thebibliography}{99}


\bibitem{AR} 
D. Allison and N. Reshetikhin, Numerical study of the 6-vertex model with domain wall
boundary conditions, {\it Ann. Inst. Fourier} (Grenoble) {\bf 55} (2005) 1847–1869.

\bibitem{BKMM} 
J. Baik, T. Kriecherbauer, K. T.-R. McLaughlin, and P.D. Miller,
Discrete orthogonal polynomials. Asymptotics and applications.
{\it Ann. Math. Studies} {\bf 164}. Princeton University Press.
Princeton and Oxford, 2007.

\bibitem{BBOY}
M.T. Batchelor, R.J. Baxter, M.J. O'Rourke, and C.M.  Yung, 
Exact solution and interfacial tension of the six-vertex model with anti-periodic boundary conditions. 
{\it J. Phys. A} {\bf 28} (1995) 2759--2770.

\bibitem {Bax}
{R. Baxter, {\it Exactly solved models in statistical mechanics,} 
Academic Press, San Diego, CA.}

\bibitem{BF}
P.M. Bleher and V.V. Fokin,
Exact solution of the six-vertex model with domain wall boundary conditions.
Disordered phase.
{\it Commun. Math. Phys.} {\bf 268} (2006), 223--284.
%
%\bibitem {BI1} 
%{P. Bleher and A. Its,  Semiclassical asymptotics of
%orthogonal polynomials, 
%Riemann-Hilbert problem, and universality in the matrix model.
%{\it Annals of Mathematics}, 1999, {\bf 150}, 185-266.}
%
%\bibitem{BI2} 
%{P. Bleher and A. Its,  Double scaling limit in the random
%matrix model: 
%the Riemann-Hilbert approach. {\it Commun. Pure Appl. Math.},
%{\bf 56} (2003), 433-516.}

\bibitem{BI3} 
{P. Bleher and A. Its,
Asymptotics of the partition function of a random matrix model.
{\it Ann. Inst. Fourier} {\bf 55} (2005), 1943--2000.}

\bibitem{BL1}
P.M. Bleher and K. Liechty,
Exact solution of the six-vertex model with domain wall boundary conditions.
Ferroelectric phase, {\it Commun. Math. Phys.} {\bf 286} (2009), 777--801.

\bibitem{BL2}
P.M. Bleher and K. Liechty,
Exact solution of the six-vertex model with domain wall boundary condition. Critical
line between ferroelectric and disordered phases, {\it J. Statist. Phys.} {\bf 134} (2009), 463--485.

\bibitem{BL3}
P.M. Bleher and K. Liechty,
Exact solution of the six-vertex model with domain wall boundary conditions.
Antiferroelectric phase. {\it Commun. Pure Appl. Math.} {\bf 63} (2010), 779--829.

\bibitem{BL4}
P.M. Bleher and K. Liechty,
Uniform Asymptotics for Discrete Orthogonal Polynomials with respect to Varying Exponential Weights on a Regular Infinite Lattice. {\it Int. Math. Res. Not.}  (2010) 0: rnq081v2-rnq081.


\bibitem{BKZ}
N.M. Bogoliubov, A.M. Kitaev, and M.B. Zvonarev,
Boundary polarization in the six-vertex model,
{\it Phys. Rev. E} {\bf 65} (2002), 026126.


%\bibitem {CP1}
%F. Colomo and A.G. Pronko,
%Square ice, alternating sign matrices, and classical orthogonal polynomials,
%{\it J. Stat. Mech. Theory Exp.} 2005, no. 1, 005, 33 pp. (electronic).


%\bibitem{CP3}
%F. Colomo and A.G. Pronko,
%The arctic circle revisited. Preprint. arXiv:0704.0362.


%\bibitem{DKM} 
%P.A. Deift, T. Kriecherbauer, K.T-R. McLaughlin,
%New results on equilibrium measure for logarithmic potentials
%in the presence of an external field.
%{\it J. Approx. Theory} {\bf 95} (1998), 388-475. 

\bibitem{DKMVZ} 
{P.A. Deift, T. Kriecherbauer, K.T-R. McLaughlin,
S. Venakides, and Z. Zhou,
 Uniform asymptotics for polynomials orthogonal with
respect to varying exponential weights and applications to
universality questions in random matrix theory.
{\it Commun. Pure Appl. Math.} {\bf 52} (1999), 1335-1425.}

\bibitem {DZ}
P. Deift and X. Zhou,
    A steepest descent method for oscillatory Riemann-Hilbert problems.
    Asymptotics for the MKdV equation,
    {\it Ann. Math.} {\bf 137} (1993), 295--368.

\bibitem{EM}
N.M. Ercolani and K.T.-R. McLaughlin,
Asymptotics of the partition function for random matrices 
via Riemann-Hilbert techniques and applications to
graphical enumeration, {\it Int. Math. Res. Not.}, {\bf 14}
(2003), 755--820.


%\bibitem{Elo}
% K. Eloranta, Diamond Ice, {\it J. Statist. Phys.} {\bf 96} (1999) 1091–1109.


\bibitem{FS} P.L. Ferrari and H. Spohn, 
Domino tilings and the six-vertex model at its free fermion point,
{\it J. Phys. A: Math. Gen.} {\bf 39} (2006) 10297–10306.

\bibitem{FIK} 
A.S. Fokas, A.R. Its, and A.V. Kitaev. The isomonodromy approach to matrix models in 2D quantum gravity. 
{\it Comm. Math. Phys.} {\bf 147} (1992), 395--430.

\bibitem{Ize}
{A. G. Izergin, Partition function of the six-vertex model in a finite
volume.
{\it Sov. Phys. Dokl.} {\bf 32} (1987), 878.}

\bibitem{ICK}
{A. G. Izergin, D. A. Coker, and V. E. Korepin,
Determinant formula for the six-vertex model.
{\it J. Phys. A}, {\bf 25} (1992), 4315.}

%\bibitem{KS}
%R. Koekoek and R. Swarttouw,
%The Askey-scheme of hypergeometric orthogonal polynomials and its $q$-analogue,
%Report 98-17, TU Delft. 

\bibitem {KS}
{R. Koekoek, P.A Lesky, R. Swarttouw, {\it  Hypergeometric orthogonal polynomials and their $q$-analogues,} 
Springer, 2010.}

\bibitem{Kor}
{ V. E. Korepin,
Calculation of norms of Bethe wave functions.
{\it Commun. Math. Phys.} {\bf 86} (1982), 391-418.}

\bibitem {KZ}
{V. Korepin and P. Zinn-Justin,
Thermodynamic limit of the six-vertex model with domain wall 
boundary conditions,
{\it J. Phys. A} {\bf 33} No. 40 (2000), 7053}

%\bibitem{KM} 
%{T. Kriecherbauer, K. T-R. McLaughlin,
% Strong asymptotics of polynomials orthogonal with
%respect to Freud weights.
%{\it Int. Math. Res. Not.} {\bf 6} (1999), 299-333.}


\bibitem{Kup}
G. Kuperberg,
Another proof of the alternating sign matrix conjecture.
{\it Int. Math. Res. Not.} (1996), 139-150.

\bibitem{Lieb1}
E. H. Lieb, Exact solution of the problem of the entropy of two-dimensional
ice. {\it Phys. Rev. Lett.} {\bf 18} (1967) 692.

\bibitem{Lieb2}
E. H. Lieb,
Exact solution of the two-dimensional Slater KDP model of an antiferroelectric.
{\it Phys. Rev. Lett.} {\bf 18} (1967) 1046-1048.

\bibitem{Lieb3}
E. H. Lieb,
Exact solution of the two-dimensional Slater KDP model of a ferroelectric.
{\it Phys. Rev. Lett.} {\bf 19} (1967) 108-110.

\bibitem{Lieb4}
E. H. Lieb,  Residual entropy of square ice.
{\it Phys. Rev.} {\bf 162} (1967) 162.

\bibitem{LW}
E. H. Lieb and  F. Y. Wu,  Two dimensional ferroelectric models,
in {\it Phase Transitions and Critical Phenomena},
C. Domb and M. Green eds., vol. 1, Academic Press (1972) 331-490.

%\bibitem{MRR1}
%W. H. Mills, D. P. Robbins, and H. Rumsey,
%Proof of the Macdonald conjecture.
%{\it Invent. Math.} {\bf 66} (1982) 73-87.

%\bibitem{MRR2}
%W. H. Mills, D. P. Robbins, and H. Rumsey,
%Alternating-sign matrices and descending plane partitions.
%{\it J. Combin. Theory, Ser. A} {\bf 34} (1983) 340-359.


%\bibitem {She}
%S. Sheffield,  Random surfaces. {\it Ast\'erisque} {\bf 304} (2005), vi+175 pp.

\bibitem{Sog}
{K. Sogo,
Toda molecule equation and quotient-difference method.
{\it Journ. Phys. Soc. Japan} {\bf 62} (1993), 1887.}


%\bibitem{SZ}
%O.F. Syljuasen and M.B. Zvonarev, Directed-loop Monte Carlo simulations of Vertex models,
%{\it Phys. Rev.} {\bf E70} (2004) 016118.


\bibitem{Sut}
B. Sutherland,
Exact solution of a two-dimensional model for
hydrogen-bonded crystals.
{\it Phys. Rev. Lett.} {\bf 19} (1967) 103-104.

\bibitem{Sze}
G. Szeg\"{o}, {\it Orthogonal Polynomials}.
Fourth edition. Colloquium Publications,
vol. 23, AMS, Providence, RI, 1975.

\bibitem{WW}
E.T. Whittaker and G.N. Watson,
{\it A course of modern analysis}. Fourth edition, reprinted. Cambridge
University Press, 2000.

\bibitem{WL}
F.Y. Wu and K.Y. Lin,
Staggered ice-rule vertex model. The Pfaffian solution.
{\it Phys. Rev. B} {\bf 12} (1975), 419--428.

%\bibitem{Ze}
%D. Zeilberger,
%Proof of the alternating sign matrix conjecture.
%{\it New York J. Math.} {\bf 2} (1996), 59-68.

\bibitem {Z-J1} 
{P. Zinn-Justin,
 Six-vertex model with domain wall boundary conditions
and one-matrix model. {\it Phys. Rev. E} {\bf 62} (2000),
3411--3418.}

%\bibitem {Z-J2} 
%P. Zinn-Justin,
%The influence of boundary conditions in the six-vertex model.
%Preprint, arXiv:cond-mat/0205192. 


\end{thebibliography}
\end{document}